\documentclass[epj]{webofc}
\usepackage[varg]{txfonts}   
\usepackage{soulutf8}
\sethlcolor{cyan}
\usepackage{graphicx}
\usepackage{caption}
\usepackage{subcaption}
\usepackage{fixltx2e}
\wocname{epj}
\woctitle{RICAP-14 The Roma International Conference on Astroparticle Physics}
\begin{document}
\title{ANAIS: Status and prospects}
%
%

\author{J. Amaré\inst{1,2} \and S. Cebrián\inst{1,2} \and C. Cuesta\inst{3} \and E. García\inst{1,2} \and C. Ginestra\inst{1,2} \and M. Martínez\inst{1,2} \and \mbox{M.A. Oliván}\inst{1,2}\fnsep\thanks{\email{maolivan@unizar.es}} \and Y. Ortigoza\inst{1,2} \and A. Ortiz de Solórzano\inst{1,2} \and C. Pobes\inst{4} \and J. Puimedón\inst{1,2} \and \mbox{M.L. Sarsa}\inst{1,2} \and J.A. Villar\inst{1,2} \and P. Villar\inst{1,2}}

\institute{ Laboratorio de Física Nuclear y Astropartículas, Universidad de Zaragoza, Pedro Cerbuna 12, 50009, Zaragoza, Spain 
\and
           Laboratorio Subterráneo de Canfranc, Paseo de los Ayerbe s/n, 22880 Canfranc Estación, Huesca, Spain
	   \and
CENPA and Department of Physics, University of Washington, Seattle, WA, USA
           \and
Instituto de Ciencia de Materiales de Aragón, Universidad de Zaragoza - CSIC
%
          }

\abstract{%
ANAIS experiment will look for dark matter annual modulation with large mass of ultra-pure NaI(Tl) scintillators at the Canfranc Underground Laboratory (LSC), aiming to confirm the DAMA/LIBRA positive signal in a model-independent way.  Two 12.5 kg each NaI(Tl) crystals provided by Alpha Spectra are currently taking data at the LSC. Present status of ANAIS detectors background and general performance is summarized; in particular, thanks to the high light collection efficiency prospects of lowering the threshold down to 1 keVee are reachable. Crystal radiopurity goals are fulfilled for \textsuperscript{232}Th and \textsuperscript{238}U chains and \textsuperscript{40}K activity, although higher than original goal, could be accepted; however, high \textsuperscript{210}Pb contamination out-of-equilibrium has been identified. More radiopure detectors are being built by Alpha Spectra. The ongoing high quantum efficiency PMT tests and muon veto characterization are also presented. Finally, the sensitivity of the experiment for the annual modulation in the WIMP signal, assuming the already achieved threshold and background in ANAIS-25 is shown. Further improvement should be achieved by reducing both threshold and background, as expected.
}
\maketitle
\section{Introduction}
\label{Intro}
ANAIS~\cite{ANAIS1,ANAIS2} (Annual modulation with NAI Scintillators) experiment will look for dark matter annual modulation with large mass of ultra-pure NaI(Tl) scintillators at the Canfranc Underground Laboratory (LSC), aiming to confirm the DAMA/LIBRA positive signal~\cite{DAMA} in a model-independent way, using the same target and technique. The experimental goals for such an experiment have been set in an energy threshold equal or below 2 keVee and a background near the threshold below 2 counts/(keVee kg day).
\section{ANAIS-25}
\label{ANAIS-25}
\begin{figure}
\begin{center}
\includegraphics[width=.5\textwidth]{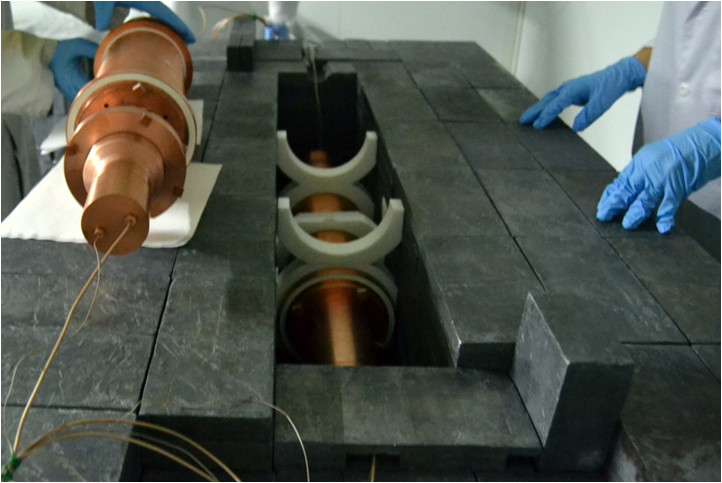}
\caption{The two modules of ANAIS-25 being introduced inside the shielding.}
\label{fig:ANAIS25}       
\end{center}
\end{figure}

ANAIS-25 prototype consists of two modules of 12.5 kg each provided by Alpha Spectra. The main goals for this prototype have been to measure the crystal internal contamination, determine light collection efficiency, fine tune the data acquisition and test the filtering and analysis protocols. The modules are cylindrical, 4.75” diameter and 11.75” length, with quartz windows for PMTs coupling. A Mylar window in the lateral face allows low energy calibration. Two types of photomultiplier have been tested: one module coupled to two Hamamatsu R12669SEL2 and the other coupled to Hamamatsu R11065SEL. The modules have been surrounded by 10 cm of archaeological lead plus 20 cm of low activity lead shielding at the Canfranc Underground Laboratory (see in Figure \ref{fig:ANAIS25}).
\begin{table}
\centering
\caption{Light collection for two ANAIS-25 modules}
\label{tab:LightCollection}       
\begin{tabular}{lll}
\hline
Detector & PMT model & phe\textsuperscript{-}/keV  \\\hline
D0 & Hamamatsu R12669SEL2 & 16.13 $\pm$ 0.66 \\
D1 & Hamamatsu R11065SEL & 12.56 $\pm$ 0.13 \\\hline
\end{tabular}
\end{table}
\begin{table}
\centering
\caption{Internal contamination measured in ANAIS-25 prototype}
\label{tab:InternalContamination}       
\begin{tabular}{cccc}
\hline
\textsuperscript{40}K (mBq/kg) & \textsuperscript{238}U (mBq/kg) & \textsuperscript{210}Pb (mBq/kg)& \textsuperscript{232}Th (mBq/kg)\\\hline
1.25 $\pm$ 0.11 (41 ppb K) &  0.010 $\pm$ 0.002& $\sim $3.15 & 0.002$\pm$ 0.001 \\
\hline
\end{tabular}
\end{table}

\begin{figure}
\begin{subfigure}[b]{0.5\textwidth}
  \begin{center}
  \includegraphics[width=1\textwidth]{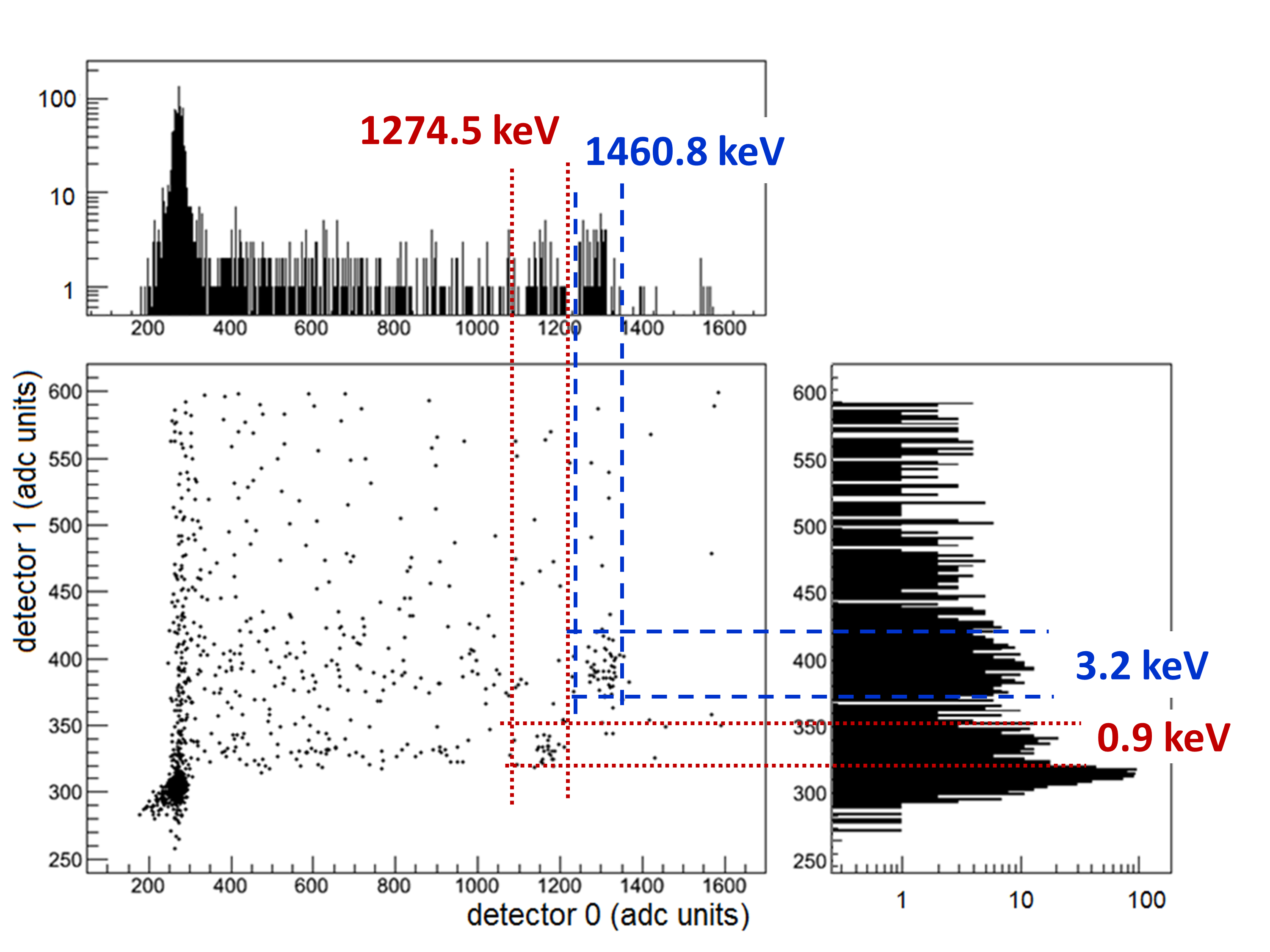}
  \caption[]{\label{fig:CoincThr}}
  \end{center}
  \end{subfigure}
        ~ 
\begin{subfigure}[b]{0.5\textwidth}
  \begin{center}
  \includegraphics[width=1\textwidth]{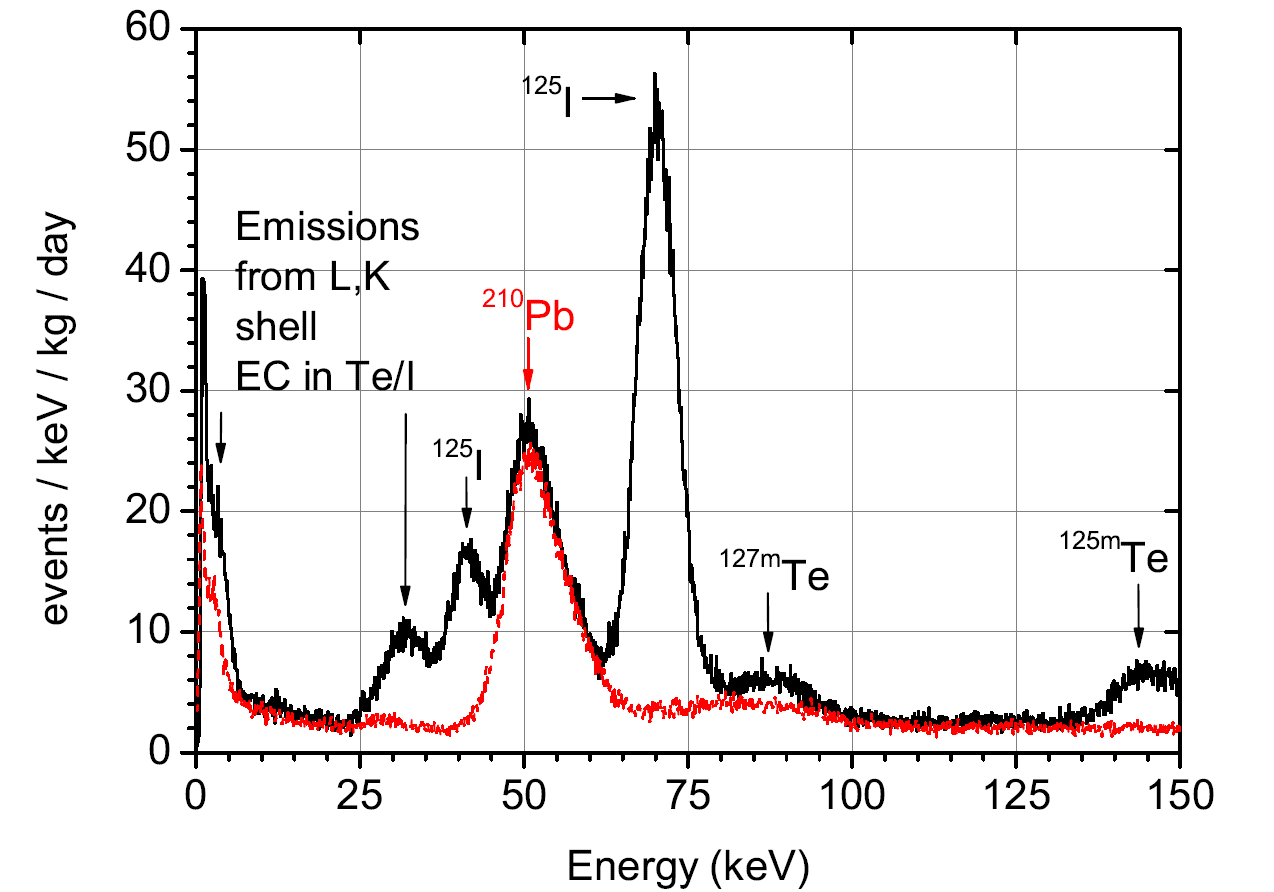}

    \caption[]{\label{fig:CosmoActiv}}
  \end{center}
  \end{subfigure}
\caption{Coincidence scatter plot showing events from \textsuperscript{40}K and \textsuperscript{22}Na (a) and cosmogenic radionuclide decay at low energy (b) showing the spectrum for the first month of data taking (black) and fifteen months later (red).}
\end{figure}

This prototype has been taking data since December 2012. The first feature to be remarked is the excellent light collection as it can be seen in Table \ref{tab:LightCollection}. This light collection has a good impact in both resolution and energy threshold. A preliminary study has been done with the coincident events in both prototype modules showing two low energy populations that have been attributed to internal \textsuperscript{40}K and cosmogenic \textsuperscript{22}Na as it is shown in Figure~\ref{fig:CoincThr}. The K-shell electron binding energy following electron capture in \textsuperscript{40}K (3.2 keV) and \textsuperscript{22}Na (0.9 keV) can be tagged by the coincidence with a high energy $\gamma$ ray (1461 keV and 1274 keV respectively). Hence, a threshold of the order of 1 keVee seems to be achievable. The trigger and filtering efficiencies at the threshold level are currently under study.

On the other hand, background contributions have been thoroughly analyzed. Figure \ref{fig:CosmoActiv} shows the low energy spectrum at the beginning of the data taking and fifteen months later, showing a high suppression of most of the lines except the corresponding to \textsuperscript{210}Pb, highlighting their cosmogenic origin; a more detailed study of radionuclide production in NaI(Tl) derived from this data can be found at reference~\cite{anais2014cosmo}. Table~\ref{tab:InternalContamination} shows the results of the activities determined for the main crystal contaminations: \textsuperscript{40}K content has been measured performing coincidence analysis between 1461 keV and 3.2 keV lines~\cite{anais2014potassium} and the activities from \textsuperscript{210}Pb and \textsuperscript{232}Th and \textsuperscript{238}U chains have been determined on the one hand, by quantifying Bi/Po sequences, and on the other, by comparing the total alpha rate with the low energy depositions attributable to \textsuperscript{210}Pb, which are fully compatible. These results give a moderate contamination of \textsuperscript{40}K, above the initial goal of ANAIS (20 ppb of K) but acceptable (see Section~\ref{sec:Prospects}), a high suppression of \textsuperscript{232}Th and \textsuperscript{238}U chains but a high activity of \textsuperscript{210}Pb at the mBq/kg level. The origin of such contamination was identified at crystal growing procedure and it is being solved by Alpha Spectra. A new module, grown following the new protocol, will be available for radiopurity checks at the beginning of 2015.

\section{Beyond ANAIS-25}
\label{BeyondANAIS-25}
The progress in the development and testing of several subsystems towards the full ANAIS experiment is reported in this section: the status of the muon veto system in Section ~\ref{sec:Veto}, the low energy calibration system in Section~\ref{sec:Calib}, the PMT quality tests in Section~\ref{sec:PMT} and other improvements in Section \ref{sec:DAQ}. 
\subsection{Muon veto}
\label{sec:Veto}
\begin{figure}
\begin{subfigure}[b]{0.5\textwidth}
  \begin{center}
  \includegraphics[width=.8\textwidth]{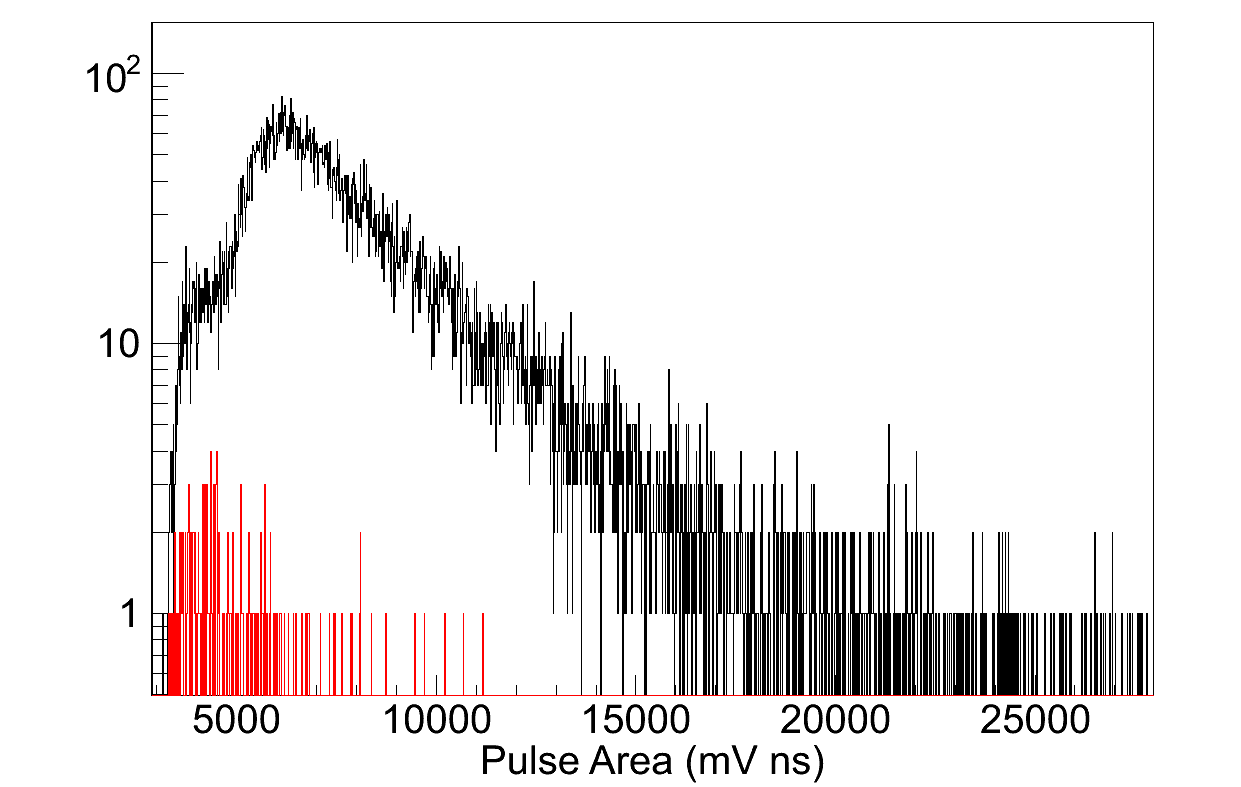}
  \end{center}
  \end{subfigure}
        ~ 
\begin{subfigure}[b]{0.5\textwidth}
  \begin{center}
  \includegraphics[width=.8\textwidth]{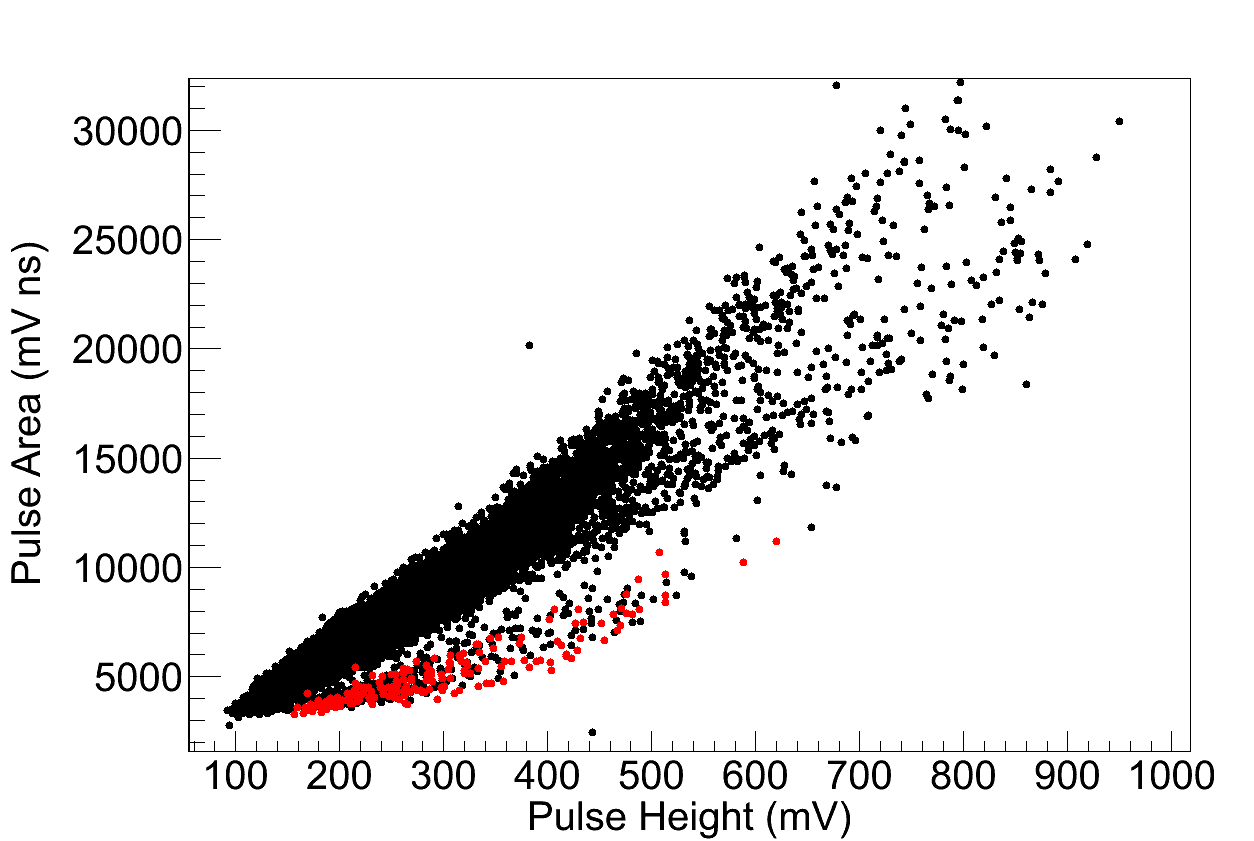}

  \end{center}
  \end{subfigure}
  \caption{Spectra (left) and scatter plot (pulse height vs. pulse area, right) taken at surface (black) and underground (red) with fast events dominating underground rate.\label{fig:MuonsAndFast}}
\end{figure}

The muon flux at the Canfranc Underground Laboratory is approximately $5 \times 10^{-3} \mu$ s\textsuperscript{-1}m\textsuperscript{-2} being one order the magnitude above the flux at LNGS.  For this reason, a plastic veto system to tag muon related events has been designed and tested in order to monitor possible systematic effects in the annual modulation analysis. The veto system will consist of sixteen plastic vetoes covering the top and all vertical faces of the shielding. The first underground data revealed a population faster than muon events and with a rate substantially higher than expected, which increased unnecessarily the experiment dead time if a \emph{threshold only} strategy is chosen to tag muons. This fast population, which is also present at surface, is clearly dominating the underground rate as can be observed in Figure~\ref{fig:MuonsAndFast}.


A dedicated front-end setup and data acquisition chain are being developed in order to separate the fast population from the real muon events. This distinction can be performed via pulse shape analysis, using shaping amplifiers and analog-to-digital converters with different temporal windows. The muon peak and the fast population are identified at surface in order to set the hardware threshold and the conditions to select muons with the different ADC values, extracting the efficiency of the cut.    
\subsection{Low energy calibration}
\label{sec:Calib}
A new calibration system scalable to a larger number of modules was developed. This system consists of flexible wires with \textsuperscript{57}Co and \textsuperscript{109}Cd sources along them. The sources are outside the shielding and they are introduced inside the shielding facing the Mylar windows during calibration keeping the system closed and Radon free. This system has been tested with ANAIS-25 modules.
%

\subsection{PMT quality tests}
\label{sec:PMT}
Forty two high quantum efficiency low background Hamamatsu R12669SEL2 PMTs have been received and are being tested. High purity germanium spectroscopy has been performed at LSC verifying the required radiopurity levels and the homogeneity among them. Operational parameters such as single electron response (SER), gain, relative quantum efficiency and dark rate are being tested at the University of Zaragoza test bench. First, a setup to expose PMTs to controlled ultraviolet LED light, data acquisition chain and algorithms to extract SER parameters have been developed and tested. The protocol to characterize all units is currently being developed. In particular, all tested units have a dark current below 500 Hz.

\subsection{Front-end, DAQ, analysis software and slow control}
\label{sec:DAQ}
The front-end setup and software chain have been redesigned in order to scale from ANAIS-25 to the full experiment. The front-end was modified with the introduction of a new high density NIM delay module and a new scalable design of linear preamplifiers. The DAQ software was designed to allow to configure the number of channels and the analysis software was modified to follow these changes. The full software stack has been tested with the low energy selection protocols from \mbox{ANAIS-0} prototype~\cite{anais2014selection} and it is being used in the following prototypes such as ANAIS-25, with configurations up to three detectors (six channels). A slow control system has also been developed and it is currently monitoring parameters such as temperatures inside the shielding and next to the electronic front-end, the external Rn activity, N\textsubscript{2} flux into the shielding and high voltage power supply voltage and current.
\subsection{Prospects}
\label{sec:Prospects}
\begin{figure}
\begin{center}
\includegraphics[width=.5\textwidth]{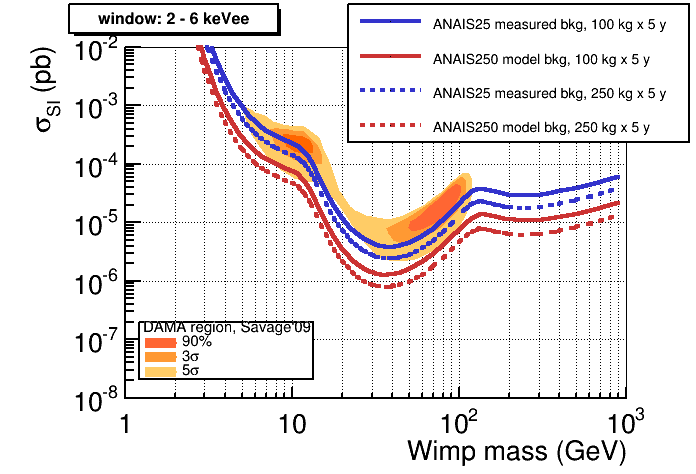}
\caption{Sensitivity projections for 2-6 keVee energy window in four different scenarios (see text).}
\label{fig:Sensitivity}       
\end{center}
\end{figure}

We have evaluated the expected sensitivity of ANAIS in different background scenarios~\cite{anais2014status}: the measured background in ANAIS-25 (in blue in Figure \ref{fig:Sensitivity}) and a model accounting for the measured contributions (PMTs, \textsuperscript{40}K and \textsuperscript{22}Na in the crystal, still without \textsuperscript{210}Pb or \textsuperscript{232}Th and \textsuperscript{238}U internal contributions, which are being added to the model) and taking into account coincidence event rejection in a 5x4 module setup (in red). The considered exposures are 100 kg x 5 years (solid lines) and 250 x 5 years (dotted). The sensitivity shown in Figure 4 corresponds to a 2-6 keVee window, and even in the most pessimistic scenario allows testing a significant part of the DAMA/LIBRA singled-out parameter space regions~\cite{savage2009}. Further sensitivity improvement should be achieved by reducing both, threshold and background, as expected.
\section*{Acknowledgments}
This work has been supported by the Spanish Ministerio de Economía y Competitividad and the European Regional Development Fund (MINECO-FEDER) (FPA2011-23749), the Consolider-Ingenio 2010 Programme under grants MULTIDARK CSD2009-00064 and CPAN CSD2007-00042, and the Gobierno de Aragón (Group in Nuclear and Astroparticle Physics, ARAID Foundation and C. Cuesta predoctoral grant). C. Ginestra and P. Villar have been supported by the MINECO Subprograma de Formación de Personal Investigador. We also acknowledge LSC and GIFNA staff for their support.
%

\begin{thebibliography}{}
%
%
\bibitem{ANAIS1}
C. Cuesta et al, Nucl. Instrum. Meth. A \textbf{742} (2014) 187-190.
\bibitem{ANAIS2}
J. Amaré et al. arXiv:1404.3564, to appear in Physics Procedia
\bibitem{DAMA}
R. Bernabei et al. Eur. Phys. J. C. \textbf{73} (2013) 2648
\bibitem{anais2014cosmo}
J. Amaré et al. arXiv:1411.0106, submitted to JCAP
\bibitem{anais2014potassium}
C. Cuesta et al. Int. J. of Mod. Phys. A. \textbf{29} (2014) 144301
\bibitem{anais2014selection}
C. Cuesta et al. Eur. Phys. J. C \textbf{74} (2014) 3150
\bibitem{anais2014status}
J. Amaré et al. arXiv:1410.5949, to appear in Nuclear Physics B - Proceedings Supplements
\bibitem{savage2009}
C. Savage et al, JCAP \textbf{04} (2009) 010

\end{thebibliography}
%
%

\end{document}